\documentclass[oribibl, 11pt]{llncs}

\usepackage[utf8]{inputenc} 
\usepackage{enumitem}
\setlist[itemize]{noitemsep, topsep=0pt} 

\usepackage{bm}

\usepackage{hyperref}

\usepackage{subcaption}

\usepackage[symbol]{footmisc}

\newcommand{\newcdot}{\cdot}

\usepackage{color}

\newcommand{\scatter}{\textsc{ScatterCache}} 
\newcommand{\nways}{$n_{ways}$} 
\newcommand{\bindices}{$b_{indices}$} 

\newcommand{\fr}{\textsc{Flush+Reload}}

\newcommand{\pp}{\textsc{Prime+Probe}} 
\newcommand{\spec}{\textsc{Spectre}} 
\newcommand{\melt}{\textsc{Meltdown}} 

\usepackage{geometry}
\usepackage{booktabs}
\usepackage{multirow}
\usepackage[table,xcdraw]{xcolor}
\usepackage{array}
\usepackage{graphicx}
\usepackage{caption} 
\newcommand{\PreserveBackslash}[1]{\let\temp=\\#1\let\\=\temp}
\newcolumntype{C}[1]{>{\PreserveBackslash\centering}p{#1}} 
\newcolumntype{?}{!{\vrule width 1pt}} 

\usepackage[most]{tcolorbox}

\newtcolorbox{profiling}{
  breakable,
  enhanced jigsaw,
  oversize,
  rightrule=0pt,
  toprule=0pt,
  bottomrule=0pt,
  colback=white,
  arc=0pt,
  outer arc=0pt,
  title={\textbf{Profiling phase.} A more effective profiling approach looks as follows.},
  title style={white},
  fonttitle=\color{black},
  titlerule=0pt,
  bottomtitle=0pt,
  top=0pt,
  bottom=0pt,
  left=10pt,
}

\newtcolorbox{covertprofiling}{
  breakable,
  enhanced jigsaw,
  oversize,
  rightrule=0pt,
  toprule=0pt,
  bottomrule=0pt,
  colback=white,
  arc=0pt,
  outer arc=0pt,
  title={\textbf{Profiling phase.}},
  title style={white},
  fonttitle=\color{black},
  titlerule=0pt,
  bottomtitle=0pt,
  top=0pt,
  bottom=0pt,
  left=10pt,
}

\newtcolorbox{coverttransmission}{
  breakable,
  enhanced jigsaw,
  oversize,
  rightrule=0pt,
  toprule=0pt,
  bottomrule=0pt,
  colback=white,
  arc=0pt,
  outer arc=0pt,
  title={\textbf{Transmitting bit $i$ of the sequence.}},
  title style={white},
  fonttitle=\color{black},
  titlerule=0pt,
  bottomtitle=0pt,
  top=0pt,
  bottom=0pt,
  left=10pt,
}

\title{Advanced profiling for probabilistic Prime+Probe attacks and
covert channels in \scatter}
\author{Antoon Purnal and Ingrid Verbauwhede}
 \institute{imec-COSIC, KU~Leuven, Belgium \\
 \texttt{\{firstname.lastname\}@esat.kuleuven.be}
 }
\date{}

\begin{document}

\maketitle

\vspace{-1.6ex}

\begin{abstract}
Timing channels in cache hierarchies are an important enabler in many microarchitectural attacks.
    \scatter~(USENIX 2019) is a protected cache architecture that randomizes the address-to-index mapping with a keyed cryptographic function, aiming to thwart the usage of cache-based timing channels in microarchitectural attacks. In this note, we advance the understanding of the security of \scatter~by outlining two attacks in the noise-free case, i.e. matching the assumptions in the original analysis. As a first contribution, we present more efficient eviction set profiling, reducing the required number of observable victim accesses (and hence profiling runtime) by several orders of magnitude. For instance, to construct a reliable eviction set in an 8-way set associative cache with 11 index bits, we relax victim access requirements from approximately $2^{25}$ to less than $2^{10}$. As a second contribution, we demonstrate covert channel profiling and transmission in probabilistic caches like \scatter. By exploiting arbitrary collisions instead of targeted ones, our approach significantly outperforms known covert channels (e.g. full-cache eviction).
\end{abstract}

\section{Introduction}%
\label{sec:introduction}

As an essential part of modern-day computing, caches hide the ever-growing latency gap between the CPU and main memory technology by exploiting locality of memory accesses. Inherent to the operation of caches, some accesses are fast and some are slow. Resulting from this fundamental timing side-channel, caches have been used as important building blocks in many micro-architectural attacks, ranging from attacks on cryptographic implementations~\cite{bernstein,osvik} to transient execution attacks like \spec~\cite{spectre} and \melt~\cite{meltdown}. Often, the term \emph{side channel} is used when an attacker uses the channel to spy on a non-cooperating victim (as in attacks on AES implementations), whereas a \emph{covert channel} denotes a channel that is used deliberately by communicating parties (as in \textsc{Meltdown}).

\textbf{Cache attacks.} Existing cache attacks can largely be classified in two categories. \emph{Removal-based} cache attacks, like \fr~\cite{flushreload} and derivatives \cite{flushflush}, can infer memory access patterns at the granularity of cache lines but require shared memory between attacker and victim. \emph{Contention-based} cache attacks like \pp~\cite{osvik,llcpractical} and \textsc{Evict+Time}~\cite{osvik} are more coarse-grained, but only rely on the shared nature of the cache. Because this enabler is readily attained in practice, contention-based attacks constitute a powerful threat. 

\textbf{Protected cache architectures.} To mitigate cache attacks, the cache hardware architecture can be augmented, giving rise to protected cache architectures. As a promising line of work, \emph{randomized} cache architectures \cite{newcache,newcache2,ceaser1,scatter} focus on randomizing the otherwise predictable mapping of memory addresses to cache sets, raising the effort for cache attacks that rely on eviction sets (like \pp). In this context, \scatter~\cite{scatter} is a recent contribution that achieves this randomization with a key-dependent cryptographic mapping. Moreover, its mapping function depends on the security domain, and separately indexes the different cache ways to increase the perceived number of cache sets. 

\textbf{Contributions}. In this note, we challenge and complement the existing security analysis of \scatter~with two main observations.
\begin{enumerate}[label=\textbf{\arabic*)}, topsep=0ex]
    \item \textbf{More efficient \pp~profiling.} The total profiling runtime is largely determined by the required number of observable victim accesses. We generalize the approach for profiling eviction sets in \scatter, significantly reducing the required victim accesses. For instance, to construct a reliable eviction set in a noise-free 8-way set associative cache with $11$ index bits, our attack accomplishes a reduction from approximately $2^{25}$ to less than $2^{10}$ victim accesses. 
    \item \textbf{Covert channels.} While \scatter~has the potential to thwart covert-channel cache attacks by significantly lowering the channel capacity, the designers do not explicitly or quantitatively consider this extended attacker model. We demonstrate a suitable approach for constructing and exploiting covert channels, yielding a covert channel that interpolates between between full~cache eviction~\cite{c5} and \pp~on traditional caches.
\end{enumerate}

\section{\scatter~preliminaries}%
\label{sec:scattercache}
Proposed at USENIX 2019, \scatter~\cite{scatter} is a promising contribution in the context of protected cache architectures. This section concisely introduces its key principles. It is not intended to present a comprehensive overview of protected cache architectures, nor does it aspire to be a complete description of \scatter. Its purpose lies in delineating the scope of this document and providing the necessary preliminaries to understand the assumptions and attacks in the sections that follow. 

As a randomized cache architecture, \scatter~replaces the predictable mapping from memory addresses to cache set indices by a pseudorandom mapping. The indexing into cache sets is performed by the Index Derivation Function (IDF). This function is instantiated with a keyed cryptographic primitive, where the key is randomly generated at system boot. 

\begin{figure}[htpb]
    \centering
    \includegraphics[width=.70\linewidth]{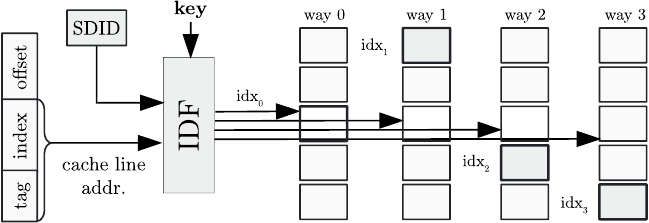}
    \captionsetup{justification=centering, margin=0.7cm}
    \caption{Cryptographic index derivation function (IDF) in \scatter~\cite{scatter}.}
    \label{fig:scattercache}
\end{figure}

Figure~\ref{fig:scattercache} presents the context in which the IDF is used. The IDF additionally considers a Security Domain Identifier (SDID) input to the mapping function, differentiating the mapping for processes belonging to other security domains. Furthermore, by separately indexing the different cache ways, \scatter~dynamically composes cache sets based on the indices in individual cache ways. As a result, the number of perceived (logical) cache sets is much larger than the physically available amount. The designers propose two constructions for the IDF: (1) The \emph{hashing} variant pseudorandomly generates output indices based on all IDF inputs; (2) The \emph{permutation} variant instantiates an individual cryptographic permutation for each of the cache ways, where the permutation is selected based on the cache tag and way index. As in the original analysis, this document focuses on the hashing variant as it is likely more secure~\cite{scatter} and straightforwardly translates to existing cryptographic primitives like (tweakable) block ciphers of conventional sizes.

In the remainder of this document, the number of ways and index bits of the cache are denoted by \nways~and \bindices, respectively.

\newpage

\section{Assumptions and simulation model}%
\label{sec:simulation_model}

In the analysis that follows, we make \textbf{identical assumptions to the original security analysis of \scatter}, which we now make explicit. Where applicable, we also describe how these assumptions are implemented in the simulator that we use to verify the attacks.

\textbf{Cryptographic unit (IDF).} As mentioned earlier, the analysis pertains to the hashing variant of \scatter. The model assumes that the mapping from memory address and SDID to output indices is perfectly pseudorandom and hence considers the cryptographic unit as a black box. Inherently, this assumption implies that there are no cryptanalytic attacks on the IDF. As a result, we assume that processes cannot, with any advantage over exhaustive search: (1) determine the memory address corresponding to a specific index; (2) determine other cache way output indices from one given output index; (3) find inputs to the cryptographic unit that produce output collisions; (4) recover the IDF key.

\textbf{Cache properties.} Faithful to how caches behave in the real world, processes cannot monitor the entries in the cache directly, nor can they infer to which cache way a certain memory address is allocated. The only interface available to them is access latency, observable by reading specific addresses; the latency is \emph{low} in case of a cache hit and \emph{high} in case of a miss. Since \scatter~flexibly adapts to a generic cache with $n_{ways}$ and $2^{b_{indices}}$ indices, the analysis is also general in these parameters. Finally, the cache has a \emph{random replacement} policy.

\textbf{Noise-free model.} Given that we match the assumptions in the security analysis of \scatter, we consider a completely noise-free model. In particular, we assume that there are no contributions of \emph{random noise} (e.g. other processes on the same machine) and \emph{systematic noise} (e.g. the code and memory of these processes do not influence the cache state). Section~\ref{sec:conclusion-future} revisits this assumption and paves the way for future research.

\textbf{Simulator.} The analysis described in the remainder of this report has been verified on a Python model of \scatter~that satisfies these assumptions. In the simulator, we instantiate the cryptographic unit with \textsc{AES}-128. Note that this does not incur a loss of generality as the attacks do not exploit any IDF internals. Moreover, in practice \scatter~would likely use a less-established cipher due to the stringent latency requirements of the CPU pipeline.

\newpage

\section{Faster \pp~profiling}%
\label{sec:profiling}

\textbf{Existing analysis.} A \pp~attack generally consist of two phases: a profiling phase and an exploitation phase, both of which are made harder by \scatter. In their paper, the \scatter~authors derive the \pp~profiling effort for the attacker in terms of number of accesses to the cache. In particular, to find $t$ addresses that collide with the victim address in at least one cache way, the expected number of victim accesses is determined as $n_{ways}^2 \cdot 2^{b_{indices}} \cdot t$. This number is very high, and prohibitively large for practical attacks. However, we show that this is a suboptimal attack strategy. Indeed, given that the profiling runtime is largely determined by the number of observable victim accesses, the profiling phase should strive to reduce these to a minimum.

\begin{profiling}

    \begin{enumerate}[label=\textbf{(\roman*)}]
        \item The attacker \textbf{generates $k$ different addresses} and reads them from memory, thereby loading them into the cache.
            \begin{itemize}
                \item[] The number of addresses $k$ is an attack parameter. Depending on $k$, there can be collisions \emph{within} the set of attacker addresses. As a key step to eliminate false positives later on, the attacker \textbf{prunes} this address set by accessing these $k$ addresses again, removing all addresses that result in high access latency (this means that they had been evicted by some other attacker address). The attacker iteratively continues the pruning this until no more addresses get evicted. Let $m_{pr}$ denote the number of pruning iterations.
            \end{itemize}
            The attacker now has a set of $k' \leq k$ addresses, which are \textbf{guaranteed to reside at a different location in the cache}. 
        \item The attacker \textbf{triggers the victim to perform the access of interest}. Specifically, the victim loads the target address, thereby evicting one of the attacker addresses with probability $p=k'/(n_{ways} \cdot 2^{b_{indices}})$. 
            \begin{itemize}
                \item[] The expected value of the cache coverage with the coupon collector problem gives an estimate of this $p$, although in practice it is lower due to the pruning.
            \end{itemize}
        \item The attacker now \textbf{accesses the set of $k'$ addresses again}, storing an address in case its access latency is high - it must have been evicted by the victim.
        \item The attacker repeats this until a set of $t$ addresses is obtained, taking on average $\frac{t}{p}$ iterations. Every iteration requires \textbf{one observable victim access} and less than $(m_{pr}+2)k$ \textbf{attacker accesses}. 
            \begin{itemize}
                \item[] Note that after the first iteration, the victim access of interest should be considered to be already in the cache. To cope with this, the attacker has two options: (1) \emph{flushing} the cache in between iterations by accessing many different addresses; (2) \emph{proceeding normally}, noting that the expected number of iterations (and hence victim accesses) increases with a factor $c \leq \min(n_{ways}, 1/p)$. Unless mentioned otherwise, we assume that attacker adopts the flushing approach to explicitly minimize the number of victim accesses.
            \end{itemize}
    \end{enumerate}
\end{profiling}

\textbf{Exploitation phase}. Resulting from the profiling phase, the attacker now has $t$ addresses that collide with the victim access of interest in at least one cache way. Proceeding with the attack, the \pp~exploitation phase is probabilistic, implying that $t$ should be chosen as a function of the desired success probability. Referring the reader to a detailed exploration of the exploitation phase in \cite{scatter}; choosing $t=275$ in an $8$-way set-associative cache with $11$ index bits results in an eviction probability of $99\%$.

\newpage

\textbf{Discussion.} We have experimentally validated the profiling procedure with the simulator described in Section~\ref{sec:simulation_model}. Profiling \pp~is much easier with this approach, notably in terms of \emph{victim accesses of interest}, which take the most effort for the attacker to obtain. Depending on the attack parameter $k$ (and hence $k'$), our procedure reduces the expected number of iterations and victim accesses $A_{v}$ from the original $A_v = n_{ways}^2 \newcdot 2^{b_{indices}} \newcdot t$ to 
\begin{equation*}
A_v = \frac{t}{p} = \frac{n_{ways} \cdot 2^{b_{indices}} \cdot t}{k'}
\end{equation*}
  (potentially multiplied with $c$). The expected number of \emph{attacker accesses} is upper-bounded by $A_a = (m_{pr}+2)kt/p$ (excluding potentially flushing the cache). Note that our profiling approach generalizes the \scatter~security analysis. The highest values for victim accesses $A_{v}$ are obtained with $k=1$, corresponding exactly to the original analysis\footnote[2]{For the $k=1$ case, it holds that $k'=1$ and $c = n_{ways}$, obtaining the expression from \cite{scatter}.}.

To provide tangible results and illustrate the scalability of the proposed approach, Table~\ref{tab:experimental_parameters} presents, for several (\nways, \bindices, $k$)-tuples, parameter values and adversarial effort to obtain one colliding address. For $t$ colliding addresses, the runtime increases linearly. Applying our approach to the AES T-tables example from the \scatter~paper~\cite{scatter} with the same assumptions (\nways $=8$; \bindices $=11$; $t=275$; cache hit $9.5ns$; cache miss $50ns$; \emph{flushing} approach between iterations which takes $3.6ms$; victim process computes $0.5ms$), the total runtime for profiling the eviction set reduces from an estimated 38 hours~\cite{scatter} to less than 5 seconds ($k=8000$). 

\begin{table}
  \centering
    \captionsetup{justification=centering, margin=0.5cm}
    \caption{As function of the attack parameter $k$: victim accesses $(A_v)$, attacker accesses per victim access $(A_a/A_v)$ and profiling runtime for one colliding address; \\ averaged over  resp. $10^7$ $(k=1)$, $10^5$ $(k=200)$ and $10^4$ $(k \geq 2000)$ simulator runs. To compute the runtime, we include the attacker miss rate $a_{miss}$.}
    \label{tab:experimental_parameters}
    \vspace{2ex}
\begin{minipage}{\linewidth}
  \centering
    \resizebox{0.8\textwidth}{!}{%
\begin{tabular}{@{}ccC{1cm}?C{0.8cm}C{0.8cm}C{1.4cm}C{1.0cm}C{1.2cm}C{1.2cm}C{1.2cm}@{}}
\toprule
    \nways & \bindices & $k$ & $m_{pr}$ & $k'$ & p & $A_{v}$ & $A_{a}/A_{v}$ & $a_{miss}$ & time \\ \midrule
    &  & 1 & 0 & 1  & $2.44 \cdot 10^{-4}$ & $4098$ & 2 & $\approx 0$ & $17$ s    \\
    &  & 200 & 2.07 & 194 & 0.047 & $21$ & 800 & $0.25$ & $86$ ms  \\
    & \multirow{-3}{*}{10} & 2000 & 4.63 & 1306 & 0.333 & $3$ & $10 \cdot 10^3$ & $0.27$ & $13$ ms \\
    & \cellcolor[HTML]{EFEFEF}{\color[HTML]{333333} } & 1 & 0 & 1 & $1.20\cdot 10^{-4}$ & $8354$ & 2 & $\approx 0$ & $34$ s  \\
    & \cellcolor[HTML]{EFEFEF}{\color[HTML]{333333} } & 200 & 1.94 & 197 & 0.024 & $42$ & 780 & 0.26 & $173$ ms\\
    \multirow{-6}{*}{4} & \multirow{-3}{*}{\cellcolor[HTML]{EFEFEF}{\color[HTML]{333333} 11}} & 4000 & 4.83 & 2610 & 0.317 & $3$ & $21\cdot 10^3$ & $0.26$ & $14$ ms \\ \bottomrule
\end{tabular}
}
\end{minipage}%
\vfill%
\vspace{2ex}%
\begin{minipage}{\linewidth}
  \centering
    \resizebox{0.8\textwidth}{!}{%
        \begin{tabular}{@{}ccC{1cm}?C{0.8cm}C{0.8cm}C{1.4cm}C{1.0cm}C{1.2cm}C{1.2cm}C{1.2cm}@{}}
\toprule
            \nways & \bindices & $k$ & $m_{pr}$ & $k'$ & p & $A_{v}$ & $A_{a}/A_{v}$ & $a_{miss}$ & time \\ \midrule
            &  & 1 & 0  & 1  & $1.23 \cdot 10^{-4}$ & $8130$ & 2 & $\approx 0$ & $33$ s\\
            & & 200 & 1.93 & 197 & 0.024 & $42$ & 780 & $0.26$  & $172$ ms\\
            & \multirow{-3}{*}{10} & 4000 & 5.12 & 2653 & $0.33$ & $3$ & $22\cdot 10^{3}$ & $0.25$ & $14$ ms\\
            & \cellcolor[HTML]{EFEFEF}{\color[HTML]{333333} } & 1 & 0 & 1 & $6.03 \cdot 10^{-5}$ & $16584$ & 2 & $\approx 0$ & $68$ s  \\
            & \cellcolor[HTML]{EFEFEF}{\color[HTML]{333333} } & 200 & 1.71 & 199 & 0.012 & $83$ & 740 & $0.27$ & $341$ ms \\
            \multirow{-6}{*}{\cellcolor[HTML]{FFFFFF}{\color[HTML]{333333} 8}} & \multirow{-3}{*}{\cellcolor[HTML]{EFEFEF}{\color[HTML]{333333} 11}} & 8000 & 5.52 & 5305 & 0.33 & $3$ & $46 \cdot 10^{3}$ & $0.23$ & $15$ ms\\ \bottomrule
\end{tabular}
    }
\end{minipage}
\end{table}

\textbf{Towards noisy environments.} Obviously, the described profiling becomes more involved when noise is present in the system. For instance, the attacker should take care that the essential pruning step effectively terminates. While investigating the attacks in the noisy case is an interesting and necessary avenue of future work, we reiterate that we simply match the assumptions in the \scatter~security analysis.

\section{Covert channels}%
\label{sec:covert_channels}

Among other microarchitectural attacks, \textsc{Meltdown}-type attacks use the cache as a covert channel. In covert channels, \emph{transmitter} and \emph{receiver} processes actively collaborate; they can already do so in the \pp~profiling phase. This allows for faster profiling than for the attack in Section~\ref{sec:profiling}, as \emph{any} collisions are now sufficient (cf. the well-known birthday problem in statistics). The correspondence with the cryptographic assumptions on the IDF is apparent: collaboration in the profiling phase reduces second-preimage search to collision search. While optimizations are possible, a baseline attack has the following steps. 

\begin{covertprofiling}
    
    \begin{enumerate}[label=\textbf{(\roman*)}]
        \item The \textbf{receiver process generates a large number of addresses}, loads them into the cache and \textbf{prunes} this address set (similar to the attacker process in Section~\ref{sec:profiling}).
        \item The \textbf{transmitter} process generates a large number of addresses and loads them into the cache.
        \item Receiver process now loads its original set of addresses again, \textbf{storing the addresses with high access latency as collision addresses}. These must have been evicted by the transmitter, indicating collision in at least one cache line.
        \item Transmitter \textbf{does the same}; it loads its original set of addresses again. Slow accesses correspond to addresses that collide with the receiver in at least one cache line.
        \item This process can be \textbf{repeated} until both transmitter and receiver \textbf{obtain a desired number of colliding addresses}. This number is an attack parameter, it can e.g. be set as a fraction $f$ of the total number cache lines (e.g. $f=0.05$)
    \end{enumerate}
    
\end{covertprofiling}

\textbf{Transmission phase.} The transmitter and receiver process now each have a set of addresses (resp. $t_T$ and $t_R$), satisfying that each transmitter address collides with at least one victim address in at least one cache way. As a result, these sets constitute a number of \emph{probabilistic covert channels}. Because each collision is assumed to occur only in one cache way, each covert channel has a $\frac{1}{n_{ways}}$ success probability. The transmitter collision addresses $t_T$ are partitioned into $s$ disjoint and equally sized bins $t_{T,i}$ (say $s=64$). 

Transmission occurs in \textbf{sequences of $s$ bits at a time}. To increase the reliability of the channel, sequences are separated by a \emph{cache flush}, performed by accessing many different addresses (excluding the $t_T$ and $t_R$ addresses), either by the transmitter or receiver process.  It is not required that the full cache be evicted; but it should flush the majority of transmitter addresses from the cache. \textbf{Within a sequence}, the transmission occurs \textbf{one bit at a time}:

\vspace{1ex}

\begin{coverttransmission}

    \begin{enumerate}[label=\textbf{(\roman*)}]
        \item The receiver loads the $t_R$ addresses into the cache.
            \begin{itemize}
                \item[] So the receiver listens on \emph{every} channel.
            \end{itemize}
        \item Depending on the value of bit $i$, the transmitter either \textbf{does} $\bm{(i=1)}$ \textbf{or does not} $\bm{(i=0)}$ access the addresses in the bin $t_{T, i}$.
            \begin{itemize}[itemsep=2ex]
                \item[] So the transmitter sends on $\frac{t_T}{s}$ channels, but sends \emph{the same bit} of information on every channel. \textbf{This redundancy overcomes the probabilistic nature of the attack}. The separation in $s$ bins ensures that not all attacker addresses are stuck in the cache after sending only one bit. Moreover, it increases the bandwidth of the attack.
            \end{itemize}
        \item The receiver now loads the $t_R$ addresses again, counting the number of slow accesses (=evictions). If this number is larger than a predetermined threshold $d$, bit $i$ is determined to be \texttt{1}, otherwise it is \texttt{0}.
            \begin{itemize}
                \item[] This threshold $d$ is determined ahead of time. Its optimal value can be greater than zero due to the non-zero probability of false positives in the channel.
            \end{itemize}
\end{enumerate}

\end{coverttransmission}

\textbf{Discussion.} We have successfully simulated both the profiling and transmission phase of the described covert channel attack, using the setup described in Section~\ref{sec:simulation_model}. The covert channel bit error rate (BER) and bandwidth depend on the cache parameters $n_{ways}$ and $b_{indices}$, and can be traded off by modifying the attack parameters $f$ and $s$. The resulting channel bandwidth interpolates a regular \pp~covert channel and a covert channel based on full-cache eviction. 

While the described covert channel profiling considers memory addresses that collide in one cache way only, finding a full cache-set collision (or anything in between) \emph{also} requires less effort if the processes collaborate. In general, the complexity of the profiling phase can be traded off with the quality of the colliding channels in the transmission phase.

\section{Conclusion and future work}%
\label{sec:conclusion-future}
Protected cache architectures constitute a promising line of work to thwart cache-based timing channels in microarchitectural attacks. In this note, we further the understanding of the residual attack surface, both for profiling side-channels and constructing covert channels. For conflict-based side-channel profiling, we generalize the existing analysis, revealing that the requirement of observable victim accesses can be lowered by several orders of magnitude. Acknowledging that \emph{completely} closing covert channels is extremely difficult, we outline an approach that has the potential to significantly outperform covert channels based on full cache evictions.

\textbf{Future work.} To strengthen the confidence in protected cache architectures and particular instances thereof, we identify interesting directions of future work:
\begin{itemize}
    \item Explore the effectiveness of these attacks in the presence of noise, both by exploring noise-reduction techniques and carefully selecting the attack parameters;
    \item Investigate formal approaches to (provably) provide lower bounds on attacker effort and upper bounds on covert channel capacities;
    \item Consider performance-preserving countermeasures (contrary to high key agility), to aid in this formal analysis and/or to limit the adversarial exposure window. 
\end{itemize}

\section*{Acknowledgements}
This work is supported in part by the Horizon 2020 research and innovation programme under grant agreement Cathedral ERC Advanced Grant 695305, and by a gift from Intel Corporation.

\newpage

\bibliographystyle{splncs04}
\bibliography{references}

\begin{thebibliography}{10}
\providecommand{\url}[1]{\texttt{#1}}
\providecommand{\urlprefix}{URL }
\providecommand{\doi}[1]{https://doi.org/#1}

\bibitem{bernstein}
Bernstein, D.J.: Cache-timing attacks on {AES}. Preprint available at
  \url{http://cr.yp.to/papers.html#cachetiming} (2005)

\bibitem{osvik}
Osvik, D.A., Shamir, A., Tromer, E.: Cache attacks and countermeasures: the
  case of {AES}. In: Cryptographers’ track at the RSA conference. pp. 1--20.
  Springer (2006)

\bibitem{spectre}
Kocher, P., Genkin, D., Gruss, D., Haas, W., Hamburg, M., Lipp, M., Mangard,
  S., Prescher, T., Schwarz, M., Yarom, Y.: Spectre attacks: Exploiting
  speculative execution. In: 2019 IEEE Symposium on Security and Privacy (2019)

\bibitem{meltdown}
Lipp, M., Schwarz, M., Gruss, D., Prescher, T., Haas, W., Fogh, A., Horn, J.,
  Mangard, S., Kocher, P., Genkin, D., Yarom, Y., Hamburg, M.: Meltdown:
  Reading kernel memory from user space. In: 27th USENIX Security Symposium.
  pp. 973--990 (2018)

\bibitem{flushreload}
Yarom, Y., Falkner, K.: Flush+ reload: a high resolution, low noise, l3 cache
  side-channel attack. In: 23rd {USENIX} Security Symposium. pp. 719--732
  (2014)

\bibitem{flushflush}
Gruss, D., Maurice, C., Wagner, K., Mangard, S.: Flush+ flush: a fast and
  stealthy cache attack. In: International Conference on Detection of
  Intrusions and Malware, and Vulnerability Assessment. pp. 279--299. Springer
  (2016)

\bibitem{llcpractical}
Liu, F., Yarom, Y., Ge, Q., Heiser, G., Lee, R.B.: Last-level cache
  side-channel attacks are practical. In: 2015 IEEE Symposium on Security and
  Privacy. pp. 605--622. IEEE (2015)

\bibitem{newcache}
Wang, Z., Lee, R.B.: New cache designs for thwarting software cache-based side
  channel attacks. ACM SIGARCH Computer Architecture News  \textbf{35}(2),
  494--505 (2007)

\bibitem{newcache2}
Wang, Z., Lee, R.B.: A novel cache architecture with enhanced performance and
  security. In: Proceedings of the 41st annual IEEE/ACM International Symposium
  on Microarchitecture. pp. 83--93. IEEE Computer Society (2008)

\bibitem{ceaser1}
Qureshi, M.K.: Ceaser: Mitigating conflict-based cache attacks via
  encrypted-address and remapping. In: 2018 51st Annual IEEE/ACM International
  Symposium on Microarchitecture (MICRO). pp. 775--787. IEEE (2018)

\bibitem{scatter}
Werner, M., Unterluggauer, T., Giner, L., Schwarz, M., Gruss, D., Mangard, S.:
  {SCATTERCACHE: Thwarting Cache Attacks via Cache Set Randomization}. In: 28th
  {USENIX} Security Symposium (2019)

\bibitem{c5}
Maurice, C., Neumann, C., Heen, O., Francillon, A.: C5: cross-cores cache
  covert channel. In: International Conference on Detection of Intrusions and
  Malware, and Vulnerability Assessment. pp. 46--64. Springer (2015)

\end{thebibliography}
    
\end{document}